# A Preference Random Walk Algorithm for Link Prediction through Mutual Influence Nodes in Complex Networks

Kamal Berahmand[1], Elahe Nasiri[2], Saman Forouzandeh[3], Yuefeng Li[4]
Department of Science and Engineering, Queensland University of Technology, Brisbane, Australia[1,4]
Department of Information Technology and Communications, Azarbaijan Shahid Madani University, Tabriz, Iran[2]
Department of Computer Engineering University of Applied Science and Technology,Tehran, Iran[3]
kamal.berahmand@hdr.qut.edu.au[1], el.nasiri@azaruniv.ac.ir[2], Saman.forouzandeh@gmail.com[3], y2.li@qut.edu.au[4]

**Abstract:**
Predicting links in complex networks has been one of the essential topics within the realm of data mining and science discovery over the past few years. This problem remains an attempt to identify future, deleted, and redundant links using the existing links in a graph. Local random walk is considered to be one of the most well-known algorithms in the category of quasi-local methods. It traverses the network using the traditional random walk with a limited number of steps, randomly selecting one adjacent node in each step among the nodes which have equal importance. Then this method uses the transition probability between node pairs to calculate the similarity between them. However, in most datasets this method is not able to perform accurately in scoring remarkably similar nodes. In the present article, an efficient method is proposed for improving local random walk by encouraging random walk to move, in every step, towards the node which has a stronger influence. Therefore, the next node is selected according to the influence of the source node. To do so, using mutual information, the concept of the asymmetric mutual influence of nodes is presented. A comparison between the proposed method and other similarity-based methods (local, quasi-local, and global) has been performed, and results have been reported for 11 real-world networks. It had a higher prediction accuracy compared with other link prediction approaches.

## 1. Introduction:

Complex networks can describe many current natural phenomena, including biological networks, brain networks, and human-made phenomena in the era of technology, such as social networks and transportation networks. This has made network science a hot, widespread, and interdisciplinary field in the current era. There are numerous problems, such as community detection [1-4], identifying spreader nodes [5, 6], maximal influence [7], and link prediction [8], in the center of these complex networks, which are considered as the main challenges. Link prediction is a critical task in complex network analysis. Link prediction approaches use past data to predict the future structure of a complex network. It can be formulated for a given social network $G$ as the prediction of the list of edges not provided in $G[t_0, t'_0]$, but predicted to exist in $G[t_1, t'_1]$, in which $G[t, t']$ implies the subgraph of $G$ at the time-stamp interval of $[t, t']$. The training interval is denoted by $[t_0, t'_0]$, and $[t_1, t'_1]$ is referred to as the testing interval. It aims at predicting missing, spurious, or new links in the current structure of the network [9]. It is a generic task for analyzing networked data, which appears in both application and theoretical analysis, new friendships, and recommend possible friends in social networks including Twitter and LinkedIn. In biological networks, it can be used to recover the consideration and understanding of protein function and discover the unknown protein-protein interactions. Link prediction in the theoretical analysis assists in comprehending the mechanism of propagation and diffusion of information [10].

Numerous link prediction algorithms have been recently proposed. Three categories of node similarity-based algorithms [11-14], maximum likelihood algorithms[15], and probabilistic models [16] are the classifications of these algorithms. In particular, similarity-based algorithms are the most effective and of the basic methods for solving the problem of link prediction. In this method, for every single pair of vertices $i$ and $j$, a score ($s_{ij}$) is calculated, which shows the similarity between the two vertices $i$ and $j$. Generally speaking, the similarity between two nodes is defined as follows: two nodes have similarities in the case of having many shared features. In terms of time and space complexity, the similarity index is classified into three categories of local, quasi-local, and global [17]. Local similarity indexes make use of structural information from vertices' neighbors to calculate their similarity, and they do not need structural information from the whole network. Compared to global similarity indexes, this approach is much faster, and it can be used in a parallel way during runtime. The main weakness of local similarity indices is that they are able to use local information from only first and second-degree neighborhoods. Most links, however, occur in paths which have more than two nodes [17]. Global similarity indexes use structural information from the whole network to score edges, and their computational complexity in large networks makes them inefficient. Also, they cannot be run in parallel mode. However, they have higher accuracy compared to local similarity indices. Quasi-local similarity indices, on the other hand, have been able to create an equilibrium between these two indexes. They use more information than local indices, and, unlike global indices, they do not make use of redundant information, which does not affect accuracy [18].

In recent years, random walk algorithms have been one of the researchers' interests because of being straightforward to interpret [19, 20]. From a practical perspective, there have been several useful applications of random walks in the area of computer science such as link prediction [21-23], community detection [24, 25], network embedding [26, 27], recommender systems [28], and diffusion on networks [29]. The graph and a starting node are given in a random walk-based method. Throughout a walk on the graph, the walker moves randomly to one of the current node's neighbors at each step. A sequence of nodes is constructed during this procedure, which determines a traverse for the graph.

Random walk is one of the primary similarity-based methods used for link prediction, which detects similarities between nodes by randomly going through the graph in global and quasi-local ways. In a global way, using the random walk with restart algorithm, the walker starts traversing from the first vertex by taking random steps and goes randomly to one of the neighbors of the first vertex with a probability of $c$, and it returns to the first vertex with a probability of $(1-c)$ [23]. The value of this index for the pair $i$ and $j$ is equal to the probability of this random walker started from vertex $i$ and locating at the vertex $j$ in the equilibrium state. This method is not very efficient for today's vast networks because of its high complexity and global information. However, the Local Random Walk (LRW) [21] algorithm limits the number of random steps to the amount $l$, and by applying this limit, the method does not have any control over equilibrium anymore. Most methods used in the random walk for the link prediction problem are using pure walking. Since in a pure random walk, the importance of all nodes and links are equally considered, the obtained result is going to be not accurate enough to recognize similarities of node pairs.

To address the above problems clearly, a modified version of the LRW algorithm is proposed. Since LRW is conducted using pure random walking and selects the destination nodes based on a random manner, its further step depends on node neighbors. Throughout its decision-making process to determine the next step, LRW randomly selects one of the neighbors using its degree. If LRW precisely selects one of the neighbors with a more significant probability, the accuracy of node similarity will improve. To help to improve the LRW, the concept of asymmetric mutual influence of nodes is presented. This concept expresses the influence of pair of nodes on each other asymmetrically. Using this concept, the walker selects the next node using its effect on the current node and selects more efficient paths for the next step. This process helps to traverse through the network structure more precisely and effectively. Therefore, nodes with a more significant structural similarity will obtain a higher score in the proposed algorithm. As a result, our proposed algorithm, called Mutual Influence Random Walk (MIRW), will be going through more efficient paths. Therefore, network structure will be examined more accurately, and more similar nodes will obtain a higher score. Compared with many other algorithms, our proposed algorithm, with its use of quasi-local information and linear time complexity, will have higher accuracy and efficiency.

The rest of this paper is organized as the following. Section 2 summarizes relevant studies on link prediction in a complex network and the existing methods for measuring the node's similarity. In Section 3, some preliminaries of the present study, including the definition of mutual information, mutual influence, preference link, and a new method of local random walk, is introduced, which depends on mutual influence for measuring the nodes' similarity in a complex network. Section 4 presents the results of experimental analysis and simulation. Finally, Section 5 provides a conclusion.

## 2. Related Works:

Recently, numerous algorithms have been implemented for link prediction, and there have already been several excellent surveys that work for the link prediction problem [9, 17, 18, 30].

Several classifications such as similarity-based algorithms, maximum likelihood methods, and probabilistic models can be provided for these methods. The maximum likelihood methods and probabilistic models provide higher accuracy than similarity-based algorithms; however, they have some intrinsic drawbacks [15]. The probabilistic models often depend on node attributes besides the network structure, so their applications are considerably restricted [12]. Furthermore, the quantity of parameters to be fixed is too large, and as a result, we cannot gain insight into the network organization, albeit building a considerably precise model. Maximum likelihood methods are not very efficient in terms of time consumption, and they can only handle the networks with hundreds of nodes [31]. In contrast, numerous real networks include nodes of different numbers from millions to billions. In this paper, we only emphasize structure-based similarity approaches using structural topology information.

The topology features of networks are applied to assign similarity scores to unconnected node pairs using structure-based similarity methods. These methods can be classified into three categories: local, quasi-local, and global [26]. Therefore, overall speaking, the similarity-based algorithms, in particular the ones based solely on quasi-local topological information, have found the widest applications. Local similarity approaches use only the information of paths with length 2 for a pair of nodes. It is divided into two main classifications, common neighbor-based and clustering coefficient-based approaches. In the category of common neighbor–based, two disconnected nodes are more probably to be mutually connected if they have more common nodes such as the Common Neighbors Index (CN) [32] directly counting the number of common neighbor nodes, Adamic–Adar Index (AA) [33] and Resource Allocation Index (RA) [13] punishing large common neighbor nodes, Sørensen Index [34], Leicht–Holme–Newman Index [12] with a penalization of large-degree endpoints.

Other approaches, such as CAR-based Common Neighbor Index (CAR), Node Clustering Coefficient (CCLP), Node and Link Clustering Coefficient (NLC), not only consider the common neighbors of node pairs but also take into account the local clustering coefficient between those common neighbors too. In the paper [35], the author considered the number of edges among the common neighbors and the CAR index presented based on the assumption that the edge exists between two nodes is more likely if their common neighbors are members of a local community (local-community-paradigm (LCP) theory). Wu et al. [36] designed the CCLP index. This index is also based on the local clustering coefficient property of the network. The local clustering coefficients of all the common neighbors of a seed node pair are computed and summed to calculate the final similarity score of the pair. The same author developed the NLC index in which combining both node and link clustering information to find the final similarity [37].

The main advantage of local similarity indices is their low computational complexity. Although, considering the immediate neighbors leads to this index to experience weak performance in prediction. On the contrary, global similarity points out the similarity according to the network's global structure information, including the Katz Index [38], counting all paths in which the connection of two nodes with shorter routes is desired. Random Walk with Restart (RWR) is a direct application of the PageRank algorithm [23]. Take a random walker into account starting from node $i$, who will iteratively move to a random neighbor with probability $c$ and come back to node $i$ with probability $(1 − c)$. Denote by $q_{ij}$ the probability this random walker locates at node $j$ in the steady-state.

Quasi-local indices do not rely on global information but they use additional topological information compared to local methods to obtain a nice trade-off between computational complexity and performance. This approach can be divided into two categories local path and random walk with finite steps. The information of all 2-step and 3-step paths, with all 2-step paths preferred, is taken into account in Local Path (LP) [39]. Effective Path (EP), Significant Path (SP), and Resources from Short Paths (RSP) [40] are the improved versions of the LP. Xuzhen et al. investigated the effective influence of endpoints and captured the connectivity, and proposed the EP in which creating the influence model among two nodes as the connectivity of paths where it is defined as the product of transfer probability of every single link included in the path [41]. Zhu et al. presented the SP index derived from the intuition that short paths make better proof of a missing link connecting its two ends (they expressed that such paths are significant); the low degree intermediate nodes are examples. Practically, the Significant Path index only applies the paths with lengths 2 and 3

[42]. Yabing et al. [40] considered the interactions of paths with different lengths based on the resource-traffic flow mechanism on networks and proposed the RSP index. Random walking with finite steps that randomly walk on the graph is very useful in calculating the similarity and proximity between nodes. The local random walk [21] and Superposed Random Walk (SRW) [21] indices are two famous random walks with finite step similarity indices. Local random walk index limiting a random walker within a local range, and superposed random walk index based on local random walk continuously releasing a random walker at the starting node to emphasize the nodes near the target node. Semi-local methods provide a trade-off between the computational complexity and the obtained accuracy. They, therefore, have been recognized as one of the most efficient approaches to deal with the link prediction problem. In semi-local methods, the local random walk algorithm is very popular and effective in finding the probability of a link existing between a pair of nodes. However, this algorithm suffers from a significant drawback in terms of accuracy. In all link prediction methods that use random walking approaches, the importance of all links and nodes is considered equal, and this makes this approach not so efficient in traversing graph structure. Here we take a different approach from previous works. In the present work, we intend to take advantage of a new concept, i.e., mutual influence, to compute the transition probability between node pairs and, therefore, not choose the random walk nodes in a purely random manner. We claim that our proposed algorithm is one of the most efficient algorithms in the semi-local category due to its high performance in the link prediction task, based on the obtained results from experiments performed on large-scale datasets.

## 3. Proposed Method:
### 3.1 Background and notation:
In this section, before getting to the algorithm, some fundamental definitions and concepts in the proposed algorithm are reviewed.
**Definition 1(Mutual Information):**
In information theory, mutual information is a concept that is a measure of the amount of information that a random variable has about another variable and also is applied to indicate the relationships between the information of nodes. Consider a couple of random variables $X$ and $Y$ with a joint probability mass function $P_{xy}$ and marginal probability mass functions $p_x$ and $p_y$ [43]. The Mutual Information $MI(X,Y)$ can be denoted as follows:

$$MI(X,Y) = \sum_{x \in X} \sum_{y \in Y} P_{xy} * \log \frac{P_{xy}}{p_x * p_y} \qquad (1)$$

$MI(X,Y)$ measures the amount of information gained by observing each of the random variables relative to the other, and has three significant features:
- MI $(X,Y)$ is always non-negative.
- MI $(X,Y)$ is zero if and only if the random variables $X$ and $Y$ are independent of each other.
- MI $(X,Y)$ = MI $(Y,X)$ In fact, mutual information is a symmetrical function.

So the above properties of $MI(X,Y)$ can measure the result of linear and nonlinear dependence between random variables $X$ and $Y$.
**Definition 2 (Asymmetric Mutual Influence(AMI)):**
In social networks, nodes have different influential and important values, and each can influence their neighbors or be influenced by their neighbors. The concept of social influence has affected various aspects of social network interactions and can be studied from different perspectives. Here, we are investigating its role in the problem of link prediction. More specifically, we take advantage of the mutual influence concept to measure how much a node can affect its neighbors and use the influence between nodes to tackle link prediction. This concept will be implemented using the network's structural information and quasi-local information of nodes. A quantity is introduced to represent the mutual influence of nodes, which uses a concept called 'Mutual Information' presented in Equation (1). We have modified this definition according to our purpose. Therefore, we measure the influence between a pair of nodes, using their first-order neighbors and the intersection of those nodes. The mutual influence between the two nodes is calculated using Equation (2.d):

$$P_i = \frac{N_i}{N} \quad (2.a)$$

$$P_j = \frac{N_j}{N} \quad (2.b)$$

$$P_{ij} = \frac{CN(i,j)}{N} \quad (2.c)$$

$$MI(i,j) = P_{ij} * \log \frac{P_{ij}}{p_i * p_j} \quad (2.d)$$

Where $N_i$ is the number of first-order neighbors of node $i$, and $N$ implies the total number of nodes in the network. $P_i$ refers to the probability of node $i$ getting influence from other nodes of the network. $CN(i,j)$ is referred to as the number of nodes direct connection to both nodes $i$ and $j$, in addition to both nodes, and $P_{ij}$ implies the occurrence probability of the intersection of node $i$ and node $j$. In fact, $P_{ij}$ is a probability that is calculated using the count of common neighbors of node $i$ and node $j$ divided by the total number of nodes in the network, and it can be interpreted as the node pair $i$ and $j$ getting influence by a set of common nodes in the network. This formula measures the mutual influence between a pair of nodes in terms of the fraction of the neighborhood that they share. Therefore, the influence that a node gives to its neighbor is equal to the influence it gets from it. But we know that in a real-life situation, this cannot be true. According to [44], the notion of influence between a couple of social entities is an asymmetric value, and it depends on various factors, e.g., an individual's importance and role in the network. We assume that the more influence a node has on its neighbor, the greater its chance to be visited from that node. Hangal et al. [44] provided a quantitative definition of influence between two entities, which is as follows:

$$Influence(i,j) = \frac{Invest(j,i)}{\sum_{k \in \Gamma_j} Invest(j,k)} \quad (3)$$

The influence that $i$ has on $j$ is determined using the amount of investment of $j$, on $i$ divided by the amount of investment of $j$ on all the other entities. The concept of investment can be interpreted as the time or effort that one person spends on the other person. In this paper, we take advantage of the concept provided by [44] and modified it to be applicable for our purpose. The new asymmetrical mutual influence, which is an asymmetrical version from Equation (2.d), is computed via the following Equation:

$$P_i = \frac{N_i}{N} \quad (4.a)$$

$$P_j = \frac{N_j}{N} \quad (4.b)$$

$$P_{ij} = \frac{P_i CN(i,j)}{\sum_{k \in \Gamma_j} CN(j,k)} \quad (4.c)$$

$$AMI(i,j) = P_{ij} * \log \frac{P_{ij}}{p_i * p_j} \quad (4.d)$$

Where $P_{ij}$, i.e., the joint probability of $i$ and $j$, is the ratio of the number of common neighbors between $i$ and $j$ to the total number of common neighbors between node $j$ and all of its neighbors multiplied by $P_i$, and $\Gamma_j$ shows the first-order neighborhood of node $j \in V$. Using this Equation means that the influence that a node gives to its neighbors depends not only on the number of common neighbors it has with that neighbor but also the number of common neighbors it has with its other neighbors. More specifically, in Equation (4.d), the maximum score is reached for nodes $i$ and $j$ when nodes $i$ and $j$ have low degrees and many mutually shared neighbors. Also, the minimum score for nodes $i$ and $j$ is reached when nodes $i$ and $j$ have high degrees and no mutually shared neighbors; under these conditions, they will be independent of each other and will not be affected by each other. Considering the following network in Figure 1 as an example, where $P_A = \frac{4}{6}$, $P_E = \frac{2}{6}$, $CN(A,E) = 3$, $\sum_{k \in \Gamma_A} CN(A,k) = 10$ and $\sum_{k \in \Gamma_E} CN(E,k) = 6$. Therefore, we can see that node E receives the strongest influence from node A, while node A receives the least influence from node E. This is happening due to the fact that node E has a lower degree compared to node A and, in addition to that, shares fewer common neighbors with its adjacent nodes compared to the number of common neighbors between node A and its adjacent nodes and, therefore, invests more resources on A, compared to the A's investments on node E.

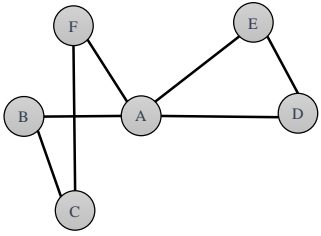
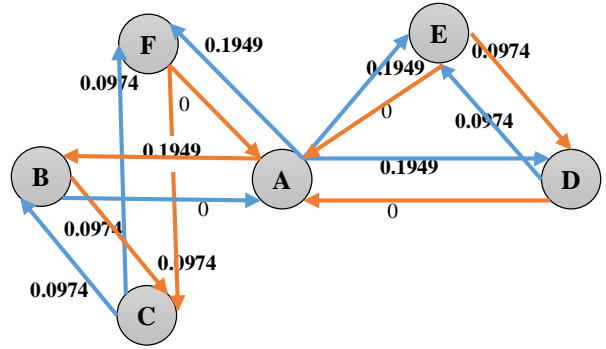

**Figure 1-A.** Before inferring Asymmetric Mutual Influence      **Figure 1-B.** After inferring Asymmetric Mutual Influence

### 3.2 Mutual Influence Random Walk (MIRW) algorithm:

Structural similarity between vertices, which are normally hidden, recognizes the similarities between nodes utilizing topological and structural information of the graph. If the structural similarity between two nodes is significant, the creation of a link between them is extremely probable. In the random-walk-based methods, if the structure of the network is traversed more efficiently, the similarity score between nodes is calculated more accurately. The local random walk and superposed random walk were some of the most effective and efficient examples of the random walk approach. They possess a significant advantage compared to other random-walk-based methods, e.g., random with restart walk, which is that these methods used quasi-local information of the network. Therefore, they significantly reduce computational complexity. The key contribution of LRW and SRW was limiting the number of steps that the walker could take. In this method, the transition probability matrix, i.e., $PR$, could be computed using the following rule:

$$PR_{ij} = \begin{cases} \frac{1}{degree(i)} & if\ (i,j) \in E \\ 0 & otherwise \end{cases} \quad (5)$$

After the transition probability matrix was obtained, the probability of the walker starting from node $i$ and reaching node $j$ after $t$ steps can be computed as follows:

$$\pi_{i,j}(t) = PR^T \pi_{i,j}(t-1) \quad (6)$$

In the above setting, $\pi_i(0)$ is a $N \times 1$ vector, with the $ith$ element equal to 1 and all the other ones equal to 0. Therefore, according to LRW, the similarity between a pair of nodes is computed using the following formula:

$$s_{ij}^{LRW}(t) = \frac{k_i}{2|E|} \cdot \pi_{i,j}(t) + \frac{k_j}{2|E|} \cdot \pi_{j,i}(t) \quad (7)$$

Where $k$ and $E$ are referred to as the degree of node and number of existing links in the network, respectively.

SRW has improved the LRW by continuously releasing walkers from the source node, resulting in a higher similarity between node pairs that were near each other. The similarity resulted from the SRW method can be obtained as follows:

$$s_{ij}^{SRW}(t) = \sum_{l=1}^{t} s_{ij}^{LRW}(l) \quad (8)$$

Even though LRW and SRW possessed many advantages like lower computational complexity and higher accuracy in predicting the missing links, compared to the previous works, their main drawback was that the process of computing the transition probability matrix was only according to the degree of the source node and therefore the results of the random walk were completely generated by random and are not accurate enough in capturing network structure and finding similarities between pairs of nodes. However, each node in the network can possess specific importance in its neighborhood and should not be treated like other nodes. To overcome this limitation, we proposed a biasing function to distinguish between every relationship of a node with its neighbors. The biasing function that is introduced in this article takes advantage of the mutual information concepts. This approach measures the influence that each node possesses on its neighbors. Therefore, the probability of the walker locating in a node, choosing one of its neighbors for the next step, is computed proportionally to the influence that it gets from that neighbor. The mutual influence between the pair nodes can be calculated by Equation (3). As previously stated, this concept is symmetrical and assumes that the influence that a node has on its neighbor is equal to the influence it gets from that neighbor. However, to produce more effectively our biasing function, we need to consider the influence of nodes on each other to be asymmetric and prefer to use AMI instead of MI. In this way, the probability of moving from node $i$ toward node $j$ is not the same as moving from node $j$ toward node $i$. Therefore, the authors use Equation (4.d) to compute a transitional probability for each pair of nodes. Consequently, we introduce a new matrix transition according to Equation (9). In this way, by tuning the parameters of the biasing function, one can force the walk to visit nodes preferentially with high values of asymmetric mutual influence.

$$PR_{ij} = PR(X_{t+1} = j | X_t = i) = \frac{P_{ij} * \log \frac{P_{ij}}{p_i * p_j}}{\sum_{j \epsilon \Gamma(i)} \left( P_{ij} * \log \frac{P_{ij}}{p_i * p_j} \right)} = \frac{AMI_{ij}}{\sum_{j \epsilon \Gamma(i)} AMI_{ij}} \quad (9)$$

Our proposed algorithm (Mutual Influence Random Walk) allows the use of an asymmetric mutual influence matrix. It is more likely to move towards a node by which it is more affected. The MIRW algorithm is defined using Equation (10):

$$S_{ij}^{MIRW}(t) = \sum_{l=1}^{t} \frac{k_i}{2|E|} \cdot \frac{AMI_{ij}}{\sum_{j \epsilon \Gamma(i)} AMI_{ij}}(l) + \frac{k_j}{2|E|} \cdot \frac{AMI_{ji}}{\sum_{j \epsilon \Gamma(i)} AMI_{ji}}(l) \quad (10)$$

By defining an appropriate weight for each pair of vertices (*i, j*), the walker jumps from one node to a neighboring node with a preference towards the link with higher weight. The pseudo-code of the proposed method is indicated below.

*Algorithm 1-The implementation procedure of MIRW similarity*

**Input:**
      $G = (V, E)$ with $n=|V|$, $m=|E|$

**Output:**
      AUC and Precision

**Begin algorithm**

1:     Divide the original network G into training set $G_{train}$ and test set $G_{test}$
2:     **For each pair of a node (i,j) in $G_{train}$ do**
3:         Compute the Asymmetric Mutual Influence (i,j)
4:     **End for**
5:     **For each unconnected pair of nodes (x,y) in $G_{train}$ do**
6:         Compute the similarity score of the edge(x,y) as $S_{xy}$ using Eq 6.
7:     **End For**
8:     Arrange the list of all $S_{xy}$ in descending order
9:     Insert top-L edges from the ordered list to $G_{train}$. //L is the number of removed edges from the original network
10:    Compute AUC and Precision
11:    **End algorithm**

## 4. Experimental analysis:

In this section, to investigate the efficiency of the proposed method, the authors have conducted some experiments and reported their results. The proposed method's performance is evaluated against some of the state-of-the-art link prediction methods. These methods were categorized according to the network's structure to local, global, and quasi-local categories. In the following sections, we describe the details of datasets used for performance analysis, compared methods, metrics for evaluation, and the results evaluations and comparisons. All the experiments were performed in a desktop pc equipped with a quad-core Intel i7 2.20GHz processor and 16GB RAM.

**4.1 Datasets:** The proposed approach is evaluated on real-world datasets. These real-word networks have some features, including the number of nodes, edges, average clustering coefficient, average shortest path, etc. A detailed description of these properties can be found in Table 1. Columns from left to right of Table 1 are respectively: network name, number of nodes ($|V|$), number of edges ($|E|$), average degree ($\langle K \rangle$), average clustering coefficient ($\langle C \rangle$), average shortest path length (ASPL), diameter (*D*). Each dataset has been collected from different domains for research and analysis purposes. Zachary Karate Club is a network consisting of 34 members of a university karate club, and each edge describes a friendship relation [45]. FOOTBALL is also a network of football games between college teams [46]. DOLPHINS is a network representing relationships between some dolphins [47]. CELEGANS is a neural network of the nematode Caenorhabditis Elegans [48]. PHYSICIANS is a network of 246 physicians being friends or trusting each other [49]. Food is a food web consisting of 128 nodes and 2075 edges [50]. SmaGri is a citation network in which nodes are documents, and a link is formed if a document is cited by another document [51]. Yeast is a network describing interactions between proteins [52]. NetScience is a co-authorship network connecting scientists [53]. King James is a network of vocabularies co-occurring in the same sentences [54]. CA-GrQc is a collaboration network covering scientific collaborations between the author's papers [55].

**Table 1. Topological details of real-world benchmark networks**

|    | Network     | \|V\| | \|E\|  | $\langle K \rangle$ | $\langle C \rangle$ | ASPL  | D  |
|----|-------------|-------|--------|---------------------|---------------------|-------|----|
| 1  | Karate      | 34    | 78     | 4.5880              | 0.588               | 2.408 | 5  |
| 2  | Football    | 115   | 613    | 10.661              | 0.403               | 2.508 | 4  |
| 3  | Dolphins    | 62    | 159    | 5.1290              | 0.303               | 3.357 | 8  |
| 4  | Celegans    | 297   | 2148   | 14.465              | 0.308               | 2.455 | 5  |
| 5  | Physicians  | 241   | 1098   | 9.1120              | 0.251               | 2.490 | 5  |
| 6  | Food        | 128   | 2075   | 32.422              | 0.335               | 1.776 | 3  |
| 7  | SmaGri      | 1024  | 4916   | 9.6020              | 0.349               | 2.981 | 6  |
| 8  | Yeast       | 2375  | 11693  | 9.8470              | 0.388               | 5.09  | 15 |
| 9  | NetScience  | 1461  | 2742   | 3.7500              | 0.878               | 5.82  | 17 |
| 10 | King James  | 1733  | 9131   | 18.500              | 0.163               | 3.38  | 8  |
| 11 | CA-GrQc     | 5242  | 14496  | 6                   | 0.529               | 7.60  | 17 |

## 4.2 The Evaluation Criteria:

For assessing the efficiency of the proposed method against compared methods, we need some evaluation metrics to measure how well each method is working. The two metrics used here are the area under the receiver operating characteristic curve (AUC) and precision. In the following subsections, we briefly introduce each metric separately, and then we describe the evaluation process.

### 4.2.1 AUC [56]:
The AUC is the most common metric for measuring how well a method distinguishes the missing link, i.e., links that will appear in the future, and non-existent edges, i.e., a pair of nodes that are not going to be connected. Almost all link prediction methods have been evaluated using this metric. In theory, this metric ranks all the non-observed links using their given score. It then counts the number of times a randomly selected missing edge is higher compared to a randomly chosen non-existent edge.

This is a time-consuming process, so in practice, when we want to evaluate a method instead of ranking all the non-observed edges, at each time, we just randomly select a missing edge and a non-existent edge and compare their scores. In n independent comparison, if n' is the number of times that the missing edge has a higher score than the non-existent edge, and n" is the number of times that both of them have the same score, then the AUC can be calculated as follows:

$$AUC = \frac{n' + 0.5n''}{n} \qquad (12)$$

If a link prediction model gives a score to non-observed links randomly, then the AUC will be equal to 0.5. So, if the resulted score is higher than 0.5, it means that the model performs better than random performance.

### 4.2.2 Precision:
The precision metric is used to measure how well the model predicts missing edges right. In other words, precision is for measuring the accuracy of the model. To measure the precision of a model, first, we need to rank all the non-observed edges using their given score in descending order. Then out of top-$L$ node pairs that have the highest score, we count the number of them that are a missing edge. Suppose $L_r$ missing edges exist in the top-L node pair. This means that the precision of the model is equal to:

$$precision = \frac{L_r}{L} \qquad (13)$$

### 4.2.3 Determination of random walk length:
According to [21], there is a positive correlation between the average shortest path distance and the appropriate length of the walk. Thus we find the best value of random walk length with respect to the average shortest path.

## 4.3 Comparison methods:
To evaluate our proposed method, we consider several baselines and state-of-the-art link methods from different categories, i.e., local, quasi-local, and global. In this section, these methods are introduced.

**Local methods:**

- **Jaccard coefficient:** this method computes the similarity of the node pair using the fraction of common neighbors they share relative to the total number of their neighbors. Jaccard coefficient for a pair of nodes can be computed as follows [57]:

$$JC(i,j) = \frac{\Gamma(i) \cap \Gamma(j)}{\Gamma(i) \cup \Gamma(j)}$$

shows the first-order neighborhood of node $i \in V. \Gamma(i)$

- **Resource allocation:** this metric also takes advantage of the concept of common neighbors to compute the similarity between a pair of nodes but penalizes the common neighbors with a higher degree. Resource allocation for a pair of nodes can be calculated as follows [13]:

$$RA(i,j) = \sum_{z \in |\Gamma(i) \cap \Gamma(j)|} \frac{1}{|\Gamma(z)|}$$

- **Adamic-Adar coefficient:** this metric works in a similar way to resource allocation, and the common neighbors with lower degrees contribute more in the similarity calculation process; however, the difference between these two methods is the way they penalize nodes with higher degrees. Adamic-Adar coefficient is computed as follows [58]:

$$AA(i,j) = \sum_{z \in |\Gamma(i) \cap \Gamma(j)|} \frac{1}{\log|\Gamma(z)|}$$

- **CCLP:** this metric also uses the common neighbors of node pairs, but instead of considering all the common neighbors equally, it assigns weights to them using the clustering coefficient of that node. CCLP for a pair of nodes is computed as follows [36]:

$$CCLP(i,j) = \sum_{z \in |\Gamma(i) \cap \Gamma(j)|} \text{Clustering Coefficient}_z$$

**Quesi-local methods:**
- **Local random walk:** this similarity index uses random walks and measures the similarity between a pair of nodes using local random walks [21]:

$$S_{i,j}^{LRW} = \frac{k_i}{2|E|} \cdot \pi_{ij}(t) + \frac{k_j}{2|E|} \cdot \pi_{ji}(t)$$

In this formulation, $\pi_{xy}(t)$ is the probability of reaching from node $x$ to node $y$ in t steps.

- **Superposed random walk:** this method works using a local random walk but gives more scores to the nodes nearby [21].

$$S_{i,j}^{SRW} = \sum_{l=1}^{t} S_{i,j}^{LRW}(l)$$

- **Local path:** this is a path-based method that uses paths with a length of 2 and 3 to compute the similarity between node pairs, but paths with a length of 2 are more important [39].

$$LP = A^2 + \alpha A^3$$

Where $A$ is the adjacency matrix.

**Global methods**
- **Random walk with restart[23]:** in this method, to find the similarity between a node and other nodes, a random walk is started from that node, and at each step, the walker decides the next node using the transition probability of edges. Also, the walker may return to the start node with the probability of α .Finally, the similarity between the start node and other nodes is determined using the probability of reaching that node.

## 4.4 Experimental results:

To evaluate our proposed method against other methods, we randomly remove 10% of edges from a dataset and consider them as missing edges. The remaining 90% of edges consist of the train set. Then we consider all the other node pairs that are not connected as non-existent edges. The union of these two sets of edges forms the non-observed edge set. After using each method to compute the score of all the non-observed edges, we evaluate the method using AUC and Precision. This process is repeated ten times for each dataset, and the average of them has been reported as final results.

Table 2 illustrates the results of our proposed algorithm and other comparing methods on eleven real-world datasets. The best AUC obtained for each dataset has been shown in highlighted in bold. It is obvious that although quasi-local methods, i.e., LRW, SRW and LP, and global methods, i.e., RWR are computationally more expensive compared to local methods, i.e., JC, RA, AA, and CCLP, they have achieved a significant advantage in results almost for all the networks. For example, in the Physicians network, global and quasi-local methods have achieved over 10% higher

AUC compared to local methods. Comparing the proposed method to the other methods, we understand that MIRW has significantly outperformed local, quasi-local, and global methods, which proves that MIRW has a huge advantage over all of them. In particular, comparing to the global method, i.e., RWR, it has been a 10%, 7%, and 11% improvement in AUC in karate, dolphins, and yeast networks, respectively, which is remarkable. Also, comparing to local methods, the performance of the proposed method was outstanding. For instance, in football, dolphins, and SmaGri networks, there has been an increase of 22%, 10%, and 8% resulted in AUC, which means that MIRW has considerably outperformed all the baseline local methods. In addition to that, comparing to quasi-local methods, the obtained results are very noticeable. To be more specific, in most of the networks, MIRW has significantly outperformed both LRW and SRW simultaneously, except for King James and NetSicence, in which the performance of MIRW was competitive. This is very important because it proves that using the concept of mutual influence in the transition probability computation process is very beneficial in the link prediction task.

Table 2-AUC results of different algorithms compared to the proposed method

| | Local | | | | Quasi-local | | | Global | Proposing |
|---|---|---|---|---|---|---|---|---|---|
| Network | JC | RA | AA | CCLP | LP | LRW | SRW | RWR | MIRW |
| Karate | 0.7464 | 0.7639 | 0.7733 | 0.8404 | 0.7898 | 0.8629 | 0.8648 | 0.8056 | **0.9057** |
| Football | 0.6443 | 0.6385 | 0.6386 | 0.8214 | 0.8472 | 0.8380 | 0.8433 | 0.8420 | **0.8603** |
| Dolphins | 0.7088 | 0.7078 | 0.7092 | 0.7460 | 0.7806 | 0.7786 | 0.7803 | 0.7363 | **0.8001** |
| Celegans | 0.8000 | 0.8767 | 0.8719 | 0.8670 | 0.8648 | 0.8666 | 0.8697 | 0.8697 | **0.8905** |
| Physicians | 0.7304 | 0.7247 | 0.7240 | 0.8529 | 0.9278 | 0.9094 | 0.8564 | 0.9256 | **0.9337** |
| Food | 0.6495 | 0.6195 | 0.6071 | 0.6323 | 0.6580 | 0.6102 | 0.6245 | 0.6103 | **0.6761** |
| SmaGri | 0.7908 | 0.8477 | 0.8432 | 0.8642 | 0.9059 | 0.9244 | 0.8676 | 0.9312 | **0.9247** |
| Yeast | 0.9116 | 0.9134 | 0.9083 | 0.9090 | 0.9560 | 0.9632 | 0.9115 | 0.8684 | **0.9705** |
| NetScience | 0.6834 | 0.6524 | 0.6424 | 0.9118 | 0.9950 | 0.9149 | 0.9924 | 0.9965 | **0.9976** |
| King James | 0.9399 | 0.9458 | 0.9234 | 0.9480 | 0.9527 | 0.9414 | 0.9843 | 0.9802 | **0.9862** |
| CA-GrQc | 0.8337 | 0.8462 | 0.8341 | 0.9263 | 0.9668 | 0.9172 | 0.9698 | 0.9698 | **0.9837** |

**4.4.2 ROC Curve:** A receiver operating characteristic curve is a graphical plot that shows how well a method identifies true positive samples and distinguishes them from negative samples. We need to plot the true-positive rate against the false-positive rate at varying thresholds to have a ROC curve. Figure 2 illustrates the ROC curves for each network and evaluates the performance of the proposed method, i.e., MIRW, against other comparing methods. The MIRW has outperformed all the methods, including local, quasi-local, and global methods, in almost all the datasets and has reached the best area under the curve. From these curves, it can be understood that using mutual influence to calculate weights of edges can greatly improve link prediction performance.

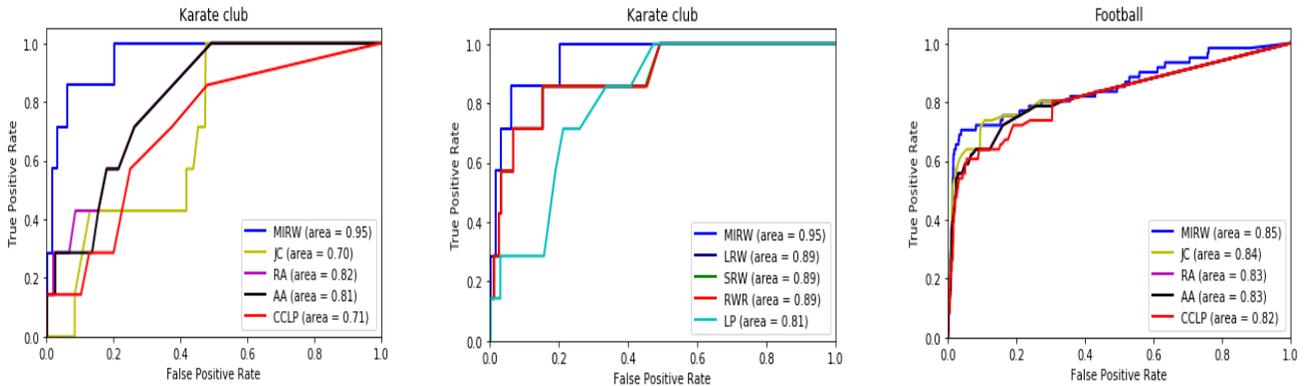

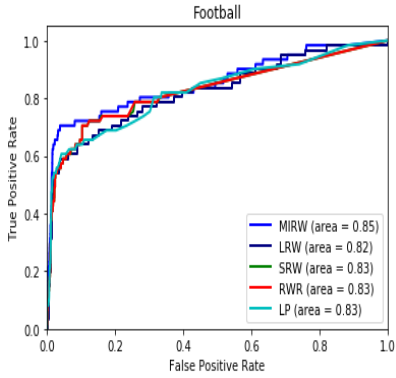
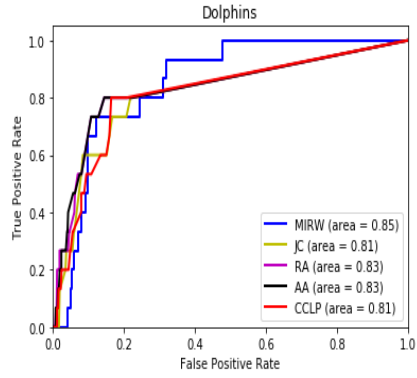
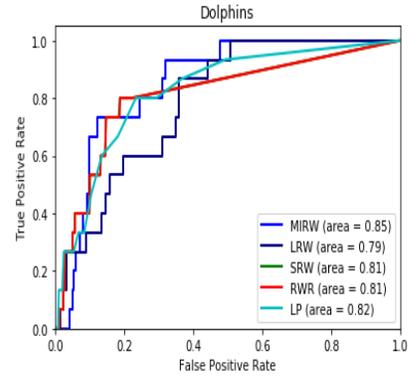
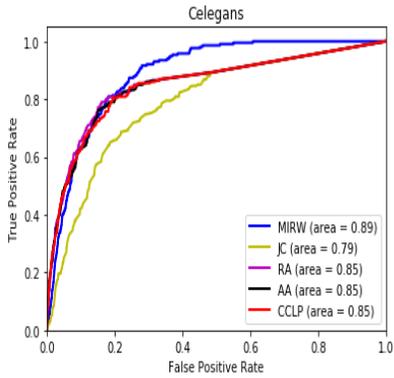
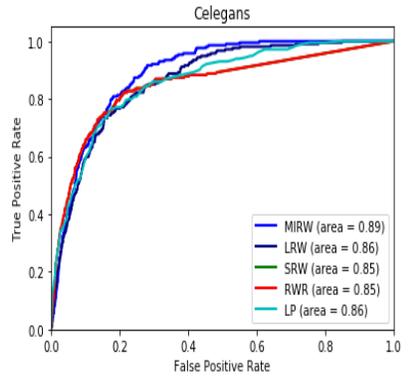
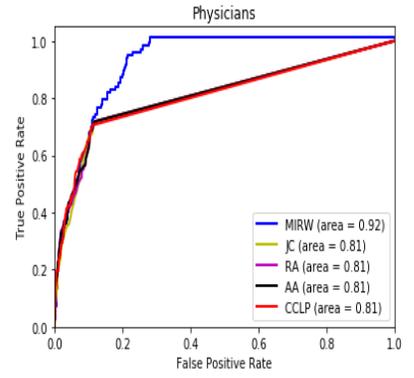
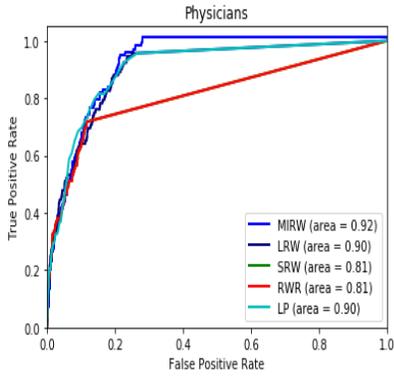
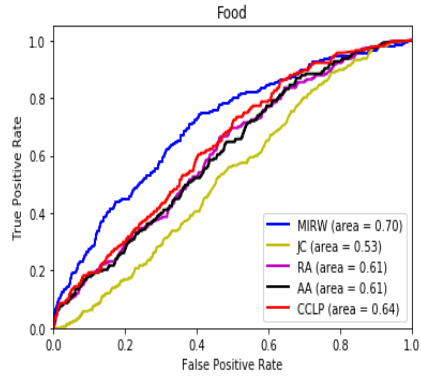
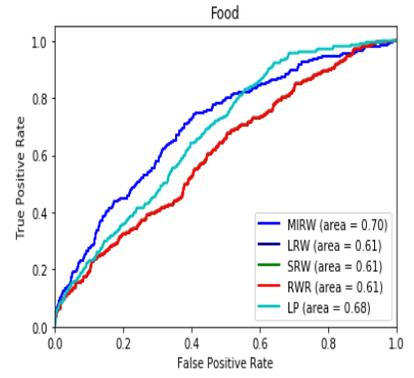
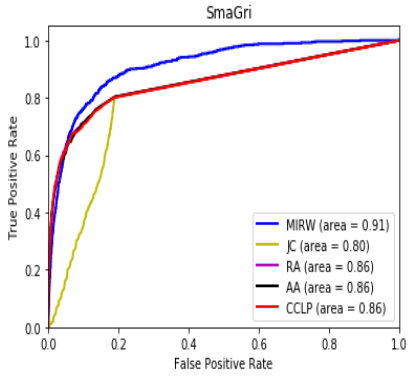
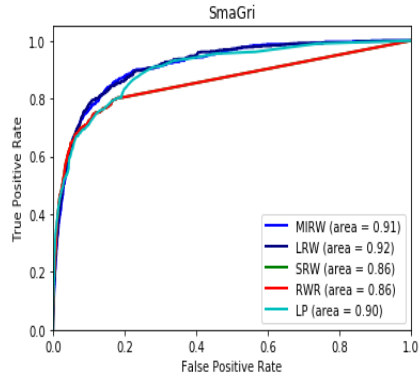
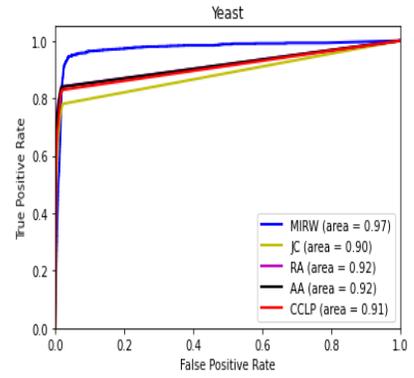

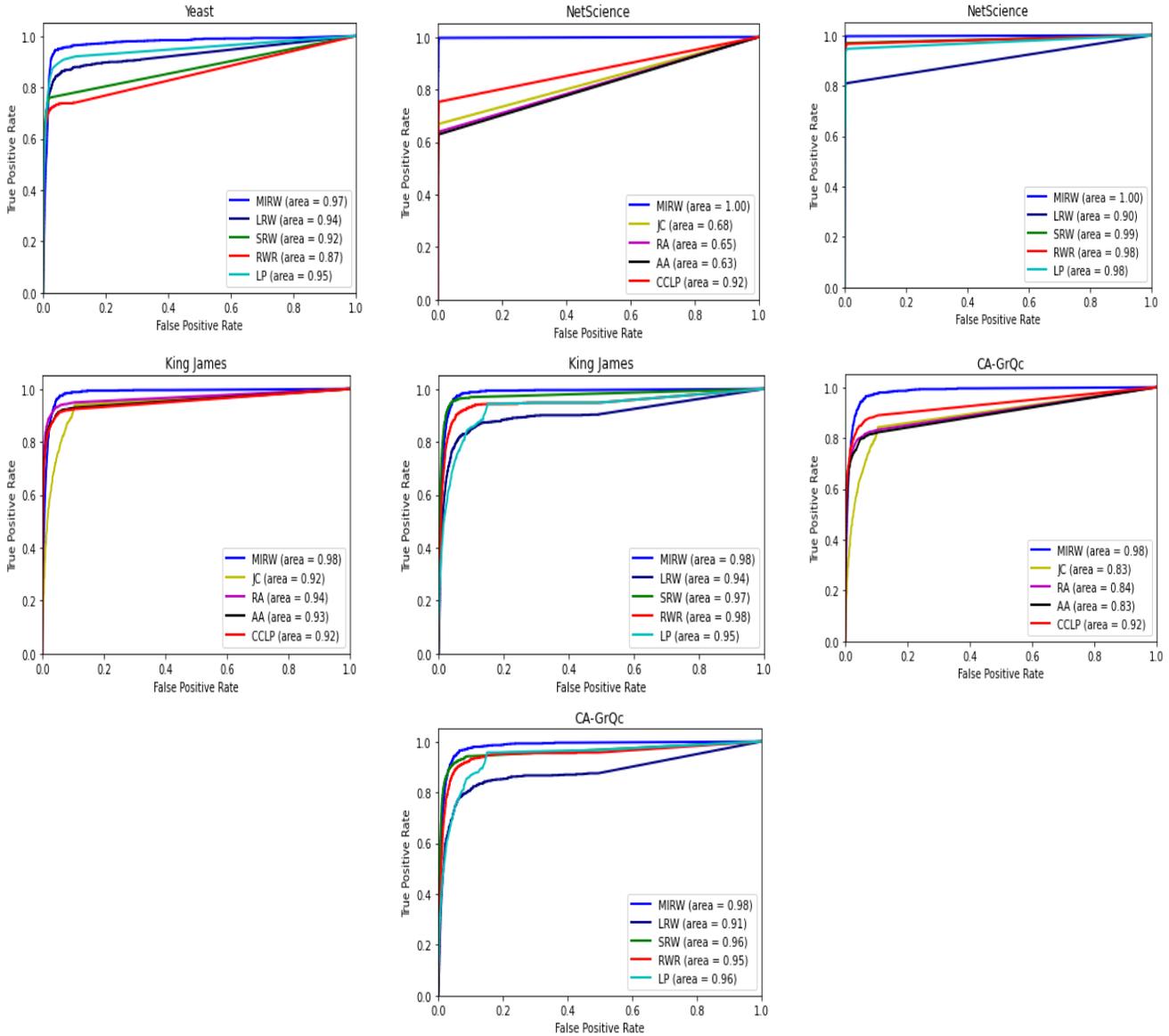

**Figure 2**- ROC Curve comparison for the proposed method vs. local methods, quasi-local and global methods

**4.4.3 Precision:** Table 3 summarizes the accuracy resulted from each method using the top precision metric. The best precision for each network is highlighted in bold. It is obvious that in most networks, our proposed method has a significant advantage compared to the local methods. In particular, in Karate, Celeganse, and Food networks, the proposed method has reached 0.3999, 0.1625, and 0.18 precision, respectively which is higher than all the local methods. However, it is clear that in some cases, the local methods have outperformed all the quasi-local and global methods in terms of top-100 precision. Compared to quasi-local and global methods, we can see that the proposed method has shown a competitive performance and proved to be precise and efficient at the same time. To be more specific, we can see that almost in all the networks, MIRW has gained more precision than LRW and SRW, which proves that taking advantage of mutual influence improves the method's performance in terms of precision. For instance, in the case of Food, Football, and King James networks, MIRW has performed 5%, 19%, and 47% better than LRW and 5%, 9%, and 36% better than SRW, which is remarkable. In other networks, MIRW has achieved acceptable results compared to LRW and SRW.

Table 3-top precision results of different algorithms compared to the proposed method

| Network | Local | | | | Quasi-local | | | Global | Proposing |
|---|---|---|---|---|---|---|---|---|---|
| | JC | RA | AA | CCLP | LP | LRW | SRW | RWR | MIRW |
| Karate | 0.0000 | 0.1999 | 0.2285 | 0.0571 | 0.1999 | 0.3141 | 0.2571 | 0.3698 | **0.3999** |
| Football | 0.3474 | 0.2917 | 0.2917 | 0.0000 | 0.2524 | 0.1999 | 0.2950 | 0.2196 | **0.3802** |
| Dolphins | 0.2034 | 0.0666 | 0.1333 | 0.0700 | 0.2033 | 0.1666 | 0.0666 | 0.1333 | **0.2077** |
| Celegans | 0.1333 | 0.0980 | 0.1366 | 0.1300 | 0.1400 | 0.1433 | 0.1566 | 0.1300 | **0.1625** |
| Physicians | 0.1195 | 0.1739 | 0.1521 | 0.1200 | 0.1195 | 0.1521 | 0.1739 | 0.1413 | **0.1883** |
| Food | 0.0400 | 0.1200 | 0.1200 | 0.1470 | 0.1370 | 0.1366 | 0.1367 | 0.1250 | **0.1800** |
| SmaGri | 0.0110 | 0.1800 | 0.1933 | 0.2166 | 0.2000 | 0.1066 | 0.1066 | 0.1166 | **0.2233** |
| Yeast | 0.5800 | 0.4900 | 0.7200 | 0.7000 | 0.6800 | 0.8600 | 0.7300 | 0.5200 | **0.8900** |
| NetScience | 0.6663 | 0.5453 | 0.5119 | 0.4800 | 0.3120 | 0.5400 | 0.5400 | 0.5500 | **0.6900** |
| King James | 0.4700 | 0.6340 | 0.5126 | **0.8300** | 0.4210 | 0.0900 | 0.2000 | 0.1400 | 0.5600 |
| CA-GrQc | 0.1200 | 0.1300 | 0.1500 | 0.1800 | **0.7800** | 0.2100 | 0.2600 | 0.2200 | 0.2700 |

**4.4.4 The varying size of the training set:** Figure 3 illustrates the effect of different training sizes on the performance of the proposed method against other methods. From this figure, it can be observed that in general, with the increase of training size, the accuracy of prediction is improving. It is obvious that almost in all datasets, the MIRW has gained higher AUC compared to local, quasi-local, and global methods in different training set sizes. This is very important because it proves that even when we have access to a small fraction of observed edges, the MIRW can still predict the non-observed edges with an impressive accuracy compared to the state-of-the-art methods. In particular, in most of the networks, when the training size is very small, the proposed method has a significant advantage over the local methods. For example, in Food, Physician, and Dolphins datasets, MIRW has 7%, 12%, and 9% higher accuracy than local methods. Compared to other quasi-local random walk based methods, i.e., LRW and SRW, the obtained results for Food, Dolphins, and Physicians networks shows approximately 6%, 5%, and 7% improvement of AUC for MIRW, which emphasizes the role of mutual influence in measuring the similarity using random walks. Also, MIRW has a noticeable advantage over the global method, i.e., RWR, when the size of the training set is very small and outperforms it in almost all the networks.

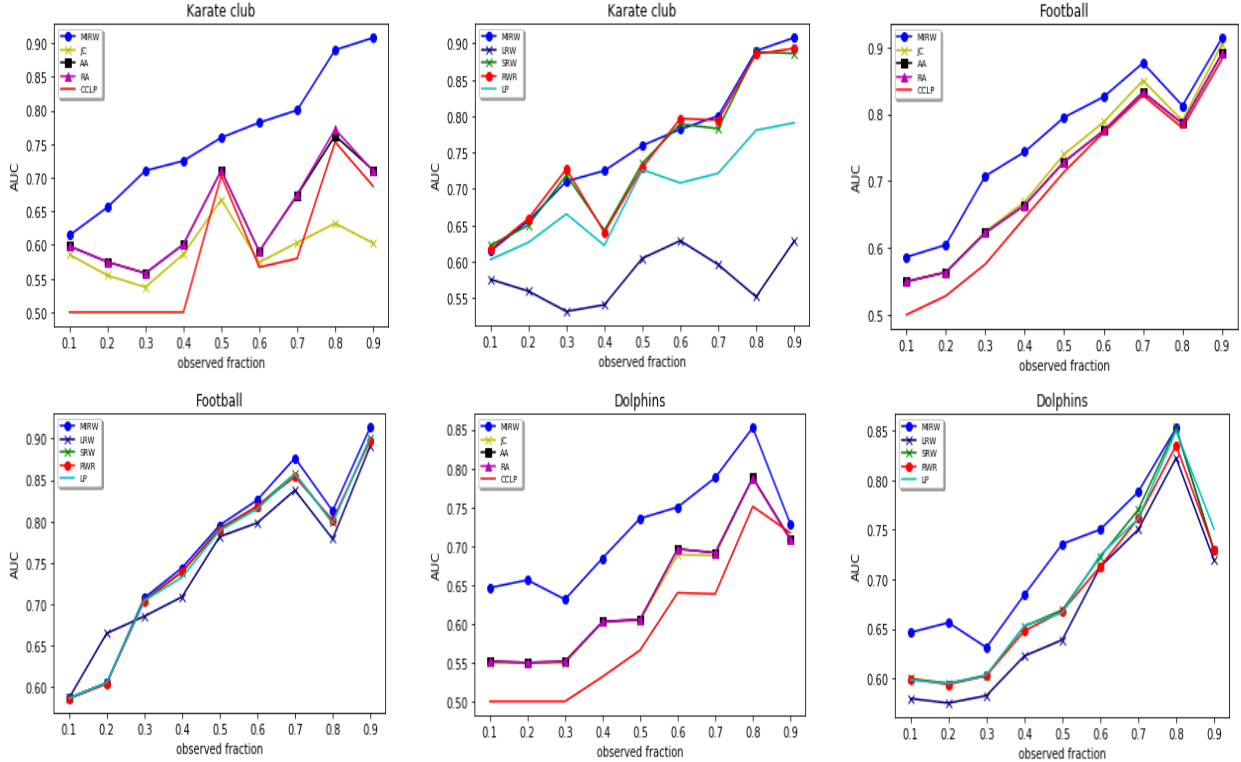

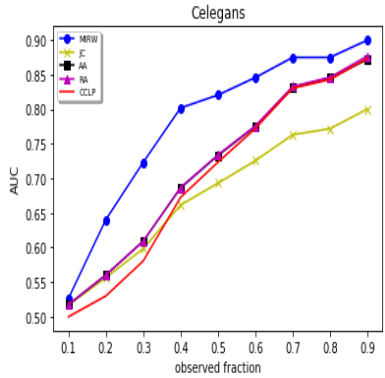
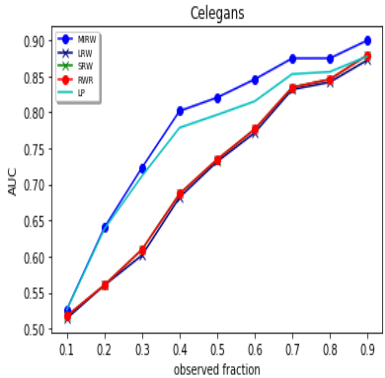
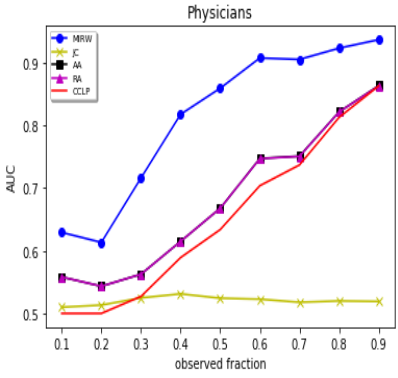
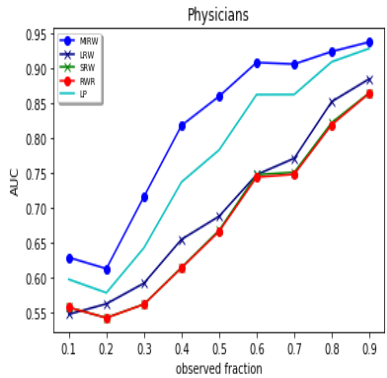
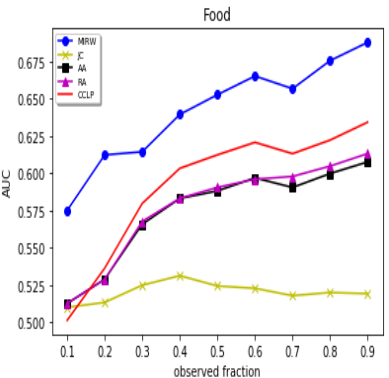
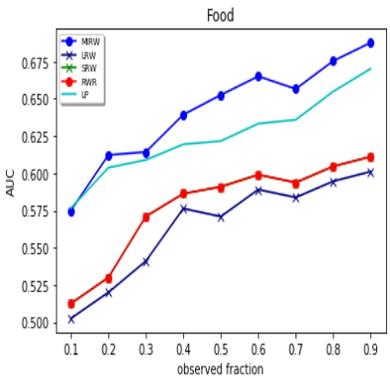
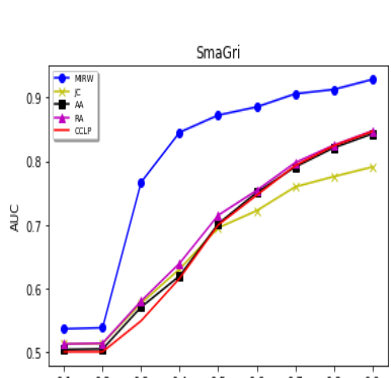
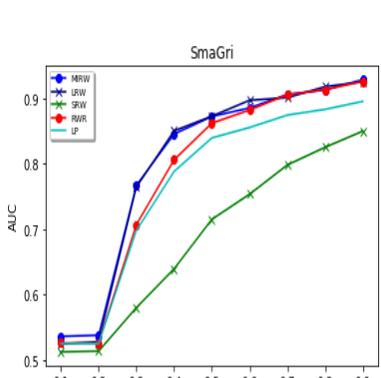
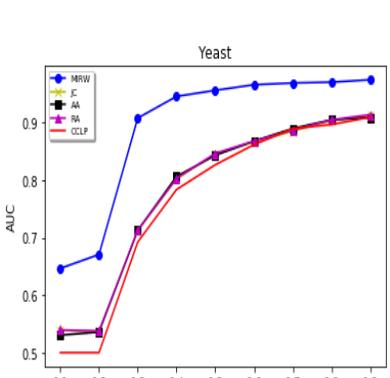
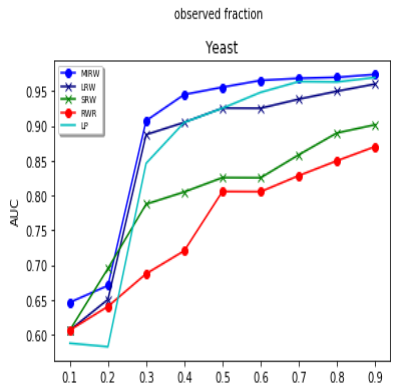
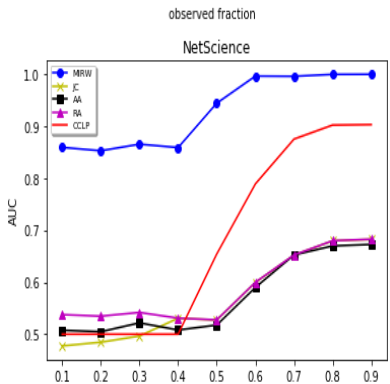
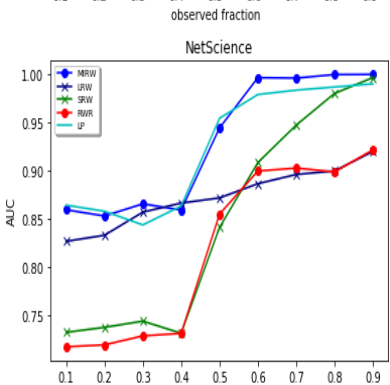

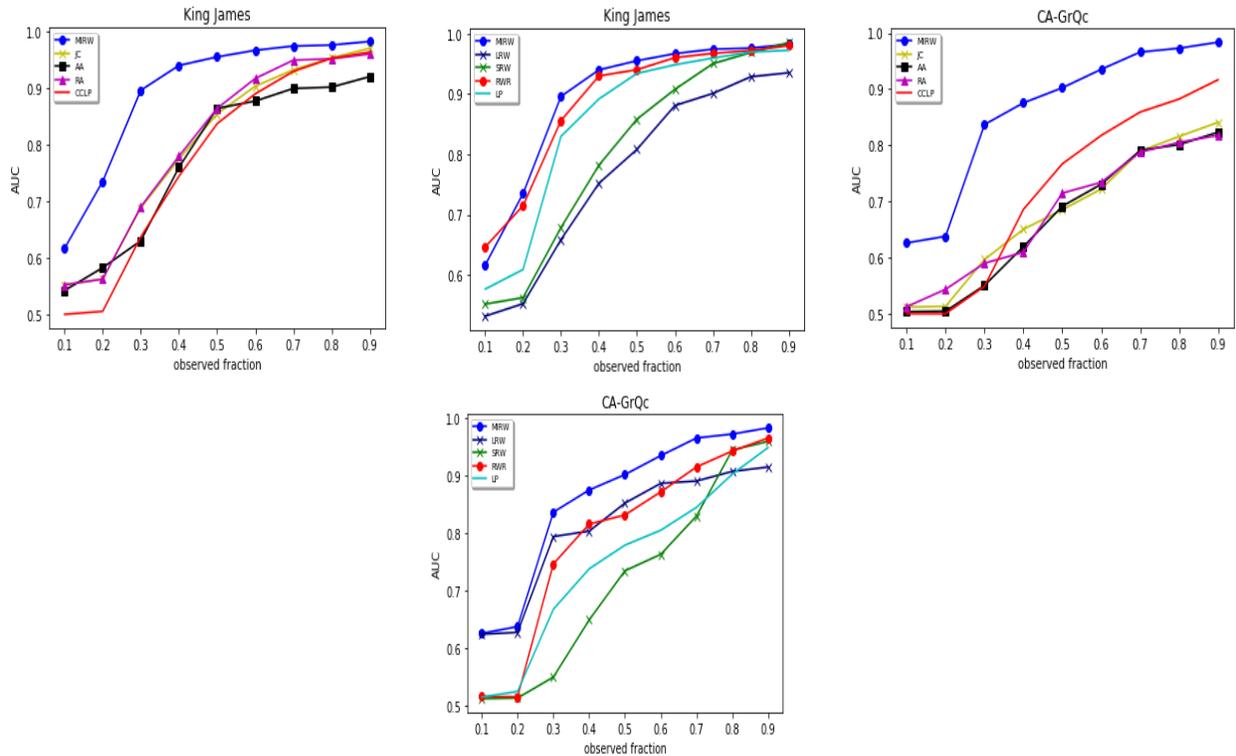

**Figure 3**- AUC comparison for the proposed method vs local methods, quasi-local and global methods

## 5. Conclusion:

In the present research, a new metric similarity is proposed for link prediction, which considers mutual influence nodes; mutual influence nodes are the interactions of two nodes between each other in an asymmetric form. Also, the proposed method takes into consideration the mutual influence neighbors of the node during the movement of the random walk to reach the next step and conducts a random walk toward the node in which the source node is affected; this results in higher efficiency compared with SRW. In order to prove the performance of our proposed approach, a comparative experiment was performed on eleven real-world networks. Our proposed approach's advantages can be observed evidently in these tests. The experimental findings from tests on many networks of various sizes indicated that the proposed plan yielded positive results than other algorithms. In future studies, the proposed method will have the option to be applied to multilayer, weighted, directed, and bipartite networks. Furthermore, suggesting an approach to specify a proper length of random walk in the proposed metric in the present study is capable of being an excellent topic for future studies.